\documentclass[manuscript]{aastex}

\usepackage{epsfig}
\usepackage{graphicx}
\usepackage{morefloats}

\shorttitle{H${\alpha}$ Profiles of Be Stars}
\shortauthors{Jones et al.}

\begin{document}

\title{The Variability of H${\alpha}$ Equivalent Widths in Be Stars}

\author{C. E. Jones\altaffilmark{1}, C. Tycner\altaffilmark{2}, A. D. Smith\altaffilmark{2}}

\altaffiltext{1}{Department of Physics and Astronomy, The University
of Western Ontario, London, Ontario, N6A 3K7, Canada}
\altaffiltext{2}{Department of Physics, Central Michigan University, Mt.
Pleasant, MI 48859, USA}

\begin{abstract}
Focusing on B-emission stars, we investigated a set of H$\alpha$ equivalent widths 
calculated from observed spectra 
acquired over a period of about 4 years from 2003 to 2007.  
During this time, 
changes in equivalent width for our program stars were monitored.  We 
have found a simple statistical method 
to quantify these changes in our observations. This statistical
test, commonly called the F ratio, involves calculating the ratio 
of the external and internal error. 
We show that the application of this technique can be used to place bounds on 
the degree of variability of Be stars.  This observational tool 
provides a quantitative way to find Be stars at particular stages of variability requiring 
relatively little observational data.
  
\end{abstract}


\section{Introduction}
\label{intro}

Be (B emission) stars are characterized by the presence of Balmer emission 
lines in their spectra due to the presence of circumstellar material.  
Often the prominent 
observational feature is the H$\alpha$ emission line, making 
this line a valuable feature to study. Accumulating evidence supports the view
that this emission originates from a geometrically thin, circumstellar disk 
rotating with near-Keplerian velocity (see, for example \citet{car06}).

Despite decades of study, the mechanism that forms and maintains these 
disks is not completely understood and this represents the main unsolved
puzzle in this field of research. Certainly, rapid rotation plays a role but 
since it is generally accepted that Be star rotation rates are below critical, 
other mechanisms must contribute to the development of the disk. 

Observations combined with detailed
modeling are crucial to interpreting disk physical conditions such as 
density, temperature, chemical composition, etc.  Constraints on
these physical properties will allow dynamic models that can follow disk structure over time to
be developed and tested with greater certainty. 

Many Be stars are known to be variable and the study of this variability has had a long history  (see the review paper by \citet{por03} for more details).  The variability occurs over a wide range of time scales from periods much less than a day
\citep{per02} to periods as long as decades \citep{oka97}. Observational evidence suggests that the early type Be stars exhibit more short-period variability compared with later types (see for example, \citet{per04, hub98}). \citet{riv03} demonstrated that the source of the short term variability is non-radial pulsation.  Long period variability has been associated with disk growth or loss events (see for example, \citet{wis10})
and cyclic changes due to disk density enhancements 
\citep{oka97}.  A successful model must account 
for this observed variability. Therefore, it is crucial 
to observationally monitor the Be star/disk system
at particular stages of variability, for example
during a disk loss or disk growth event, if we hope to improve our 
understanding of these objects. 

We have applied a statistical method to our observations in the H$\alpha$ emitting region
of Be stars to find and to classify particular systems based on their 
degree of variability. This method utilizes the ratio of 
external to internal error (E/I) and number of degrees of freedom of the system.
This test is commonly known as the E/I test, F test or variance ratio. 
Please see \citet{fis87} for more detail. Our application of the F test uses the equivalent width ($EW$) 
of the H$\alpha$ spectral line to place a quantitative bound 
on the degree of variability of a 
particular star/disk system and in doing so, allows one to 
determine if a particular system should be monitored more frequently.
Basically, this method tests the null hypothesis so if the errors are equal
then the ratio will be 1.  In our case, a rejection of the null hypothesis
means that the system is variable. We note that the F test has been used 
previously in astrophysics.  For example, \citet{gar80} 
used this test to search for binary systems in O stars.

This paper is organized as follows; the observations and data analysis can be
found in Section 2, and the results in Section 3.  A
discussion including comparisons with previous studies is provided in
Section 4.  Section 5 summarizes our work.

\section{Observations and Data Analysis}

Spectroscopic observations of the 56 program stars were acquired between 2003
December and 2007 December at the Lowell Observatory's 42-inch John
S. Hall telescope (located near Flagstaff, Arizona) equipped with the 
Solar Stellar
Spectrograph (SSS). The SSS instrument is an echelle 
spectrograph with a resolving power of 10,000 in the H$\alpha$ region.
The spectroscopic observations were processed using routines developed for 
the SSS instrument \citep{hal94}. 
There are a number of observations not
included in the analysis where the flux at the emission line has been 
saturated due to an overexposure.

\begin{figure}
\plotone{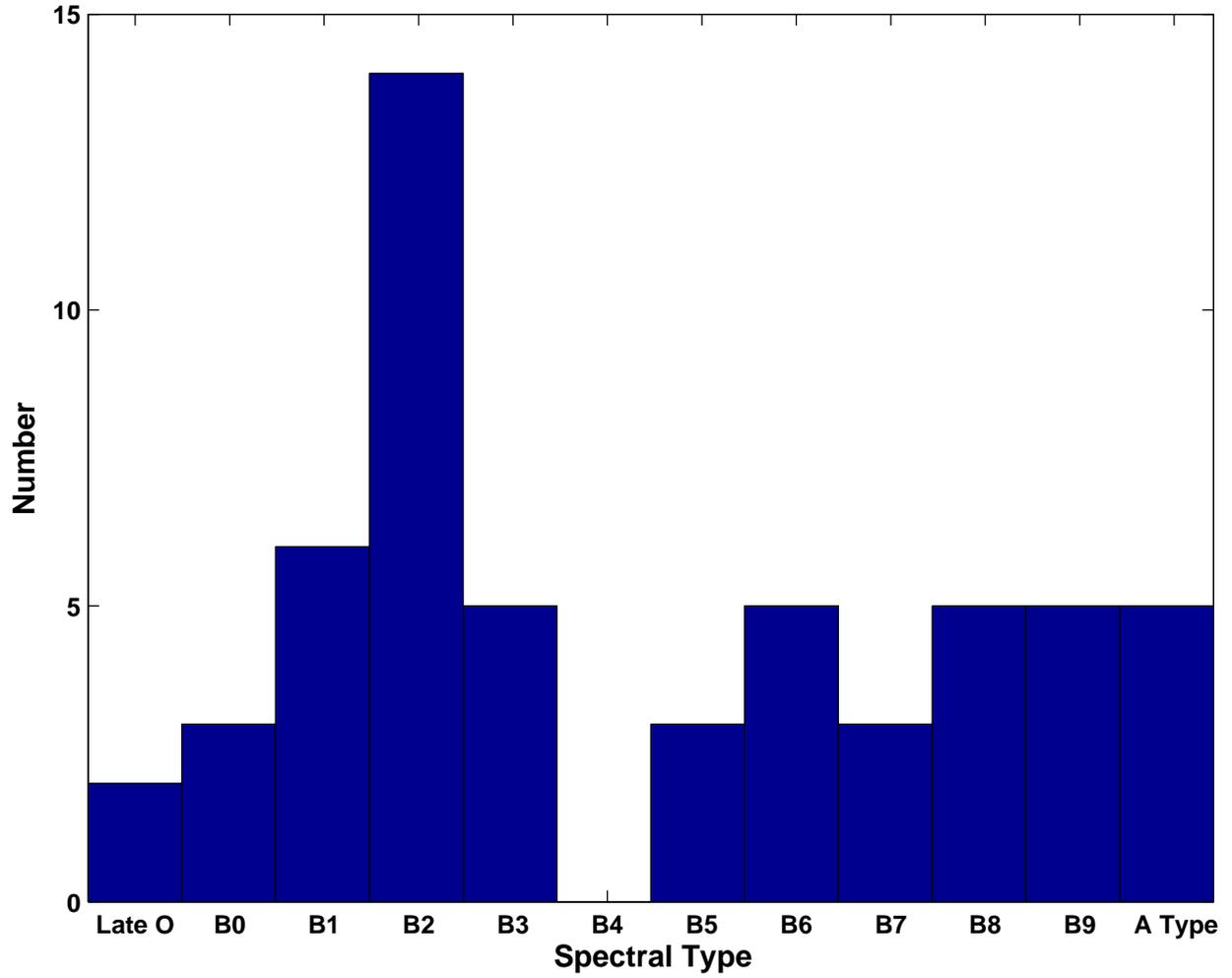}
\caption{A histogram showing the distribution of program stars as a function of spectral type.
}
\label{type}
\end{figure}

Two criteria were used to select the program stars; known Be stars that were brighter than approximately magnitude 5 in the V band, and those with declinations north of about $-20^o$ to be accessible from the telescope at air-masses of less than 2. However, in order to include as many Be stars as possible, there were some exceptions.  For example we included fainter Be stars with strong H$\alpha$ emission and bright Be stars with declinations below the threshold. We note that 49 of the 56 program stars in this
study are B spectral types, however five A-type stars and two late O-type
stars that showed H$\alpha$ emission were also included in the study. The
full list of targets can be found in Table~\ref{table_summary}. Figure~\ref{type} shows the distribution of the program stars as a
function of spectral type. Due to the difficulty of assigning spectral types for Be stars, there is likely considerable error in the spectral designation (and luminosity class) of particular stars (see \citet{ste99} for further discussion). We note that there are significantly more B2 stars compared with other types in the study. In fact, the B2 spectral types represent 25\% of the program stars. This distribution of Be stars is typical of other studies. For example, figure 1 in \citet{por96} and figure 5 in \citet{sle82} show this characteristic maximum for Be stars at spectral type B2. 

\begin{figure}
\plotone{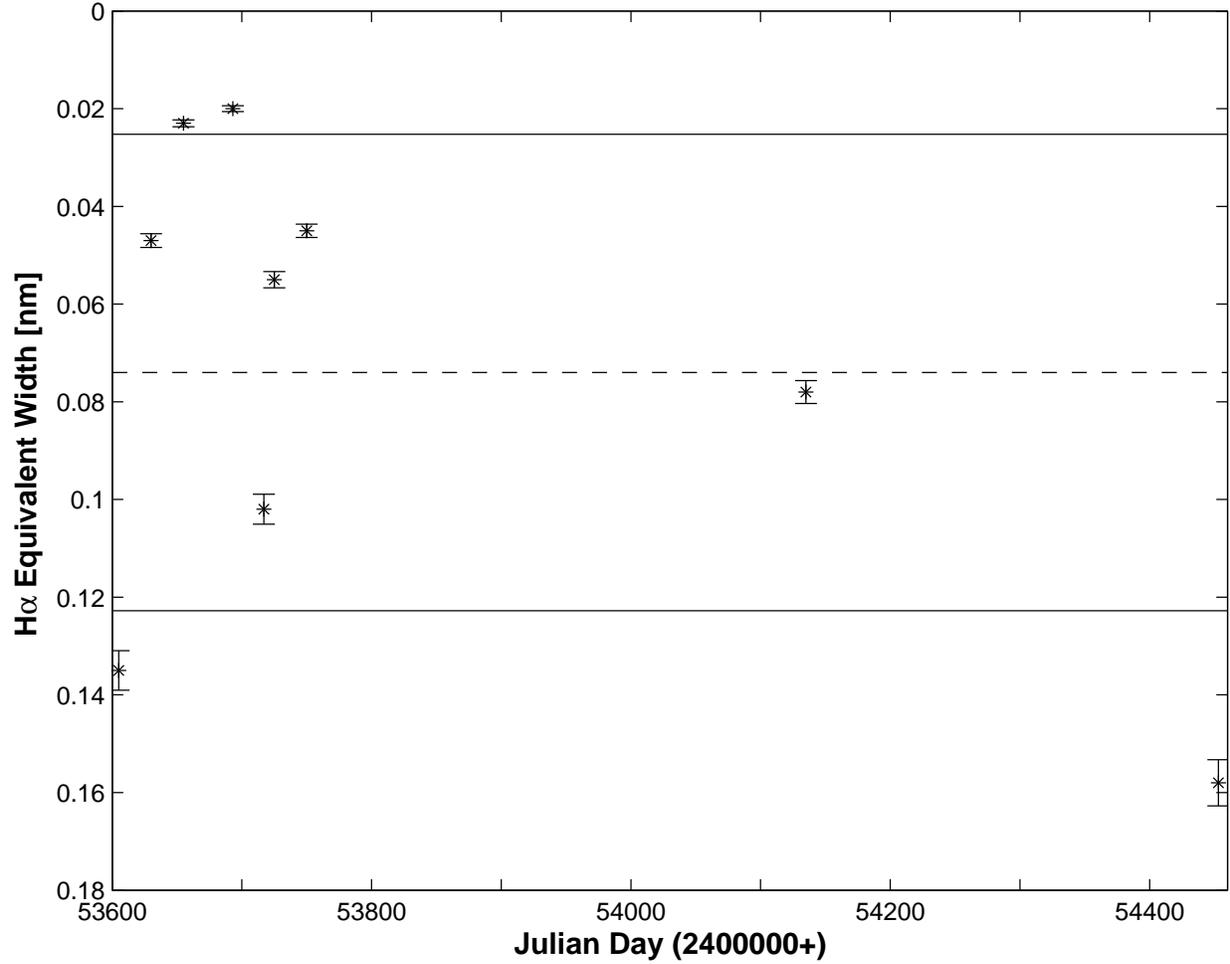}
\caption{A representative plot showing the change in H$\alpha$ equivalent 
width as a function of time for the star 23 Tau (HR 1156). The estimated error for each observation is shown by the errors bars. The 
middle dashed horizontal line
corresponds to the mean of the $EW$.  The upper and lower solid horizontal lines correspond to $\overline{EW}$ $\pm$ one standard deviation.
}
\label{example}
\end{figure}

For each target, a plot showing the variation of H$\alpha$ equivalent
width ($EW$) with time including the estimated error was
constructed. An example of such a plot is shown in
Figure~\ref{example} for the star 23 Tau (HR 1156).  Each observation
is shown along with its estimated error.  Also shown in the plot is
the mean $EW$, and the mean $\pm \,$ one standard deviation calculated
based on all the samples.  Throughout this paper, we follow the
standard convention of denoting the $EW$ of the H$\alpha$ line as
positive when observed in absorption and negative when in emission.

Technically the uncertainty of the $EW$ is based on two sources of
error; continuum normalization and the combination of photometric noise 
in the presence of telluric lines. The first and most dominant being the error associated with
continuum normalization.  Because the $EW$ is measured with respect to
the continuum, any scaling error in the level of the continuum
translates directly to a multiplicative error in $EW$.  Based on our
experience on how well the continuum normalization can be performed, and
how reproducible it is from exposure to exposure and from star to star, we
estimate this source of error to be at the 3\% level. 

The second source of error in $EW$ is related to the combination of
photometric noise (which is low for most of our bright sources) and
the presence of telluric lines.  The telluric lines vary with the
amount of atmospheric absorption, which not only changes during the
night, but also from night to night (especially across different
seasons) as the atmospheric conditions vary, mostly due to changes
in the amount of water vapor. For example, we used spectra of the telluric standard $\alpha$ Leo (HR 3982) to estimate the average impact of telluric lines across the H$\alpha$ line. We analyzed ten spectra of $\alpha$ Leo obtained on different nights and varying airmasses that were comparable to the typical conditions at which spectra of Be stars were obtained. We then fitted a smoothed Voigt profile to the absorption line to estimate the telluric component that contributed to the overall equivalent width of the H$\alpha$ line. The telluric component in the H$\alpha$ line of $\alpha$ Leo (with an $EW$ of 0.60 nm) ranged from 0.015 to 0.025 nm, which represents a fractional contribution of 2\% to 4\%.  Therefore, we note that by ignoring the telluric contribution we might be underestimating the strength of the H$\alpha$ emission in our program stars by as much as 4\%, and we are not accounting for any possible additional variability that might be present due to changes in telluric line absorption at the $\sim$2\% level. 

Our tests to establish the telluric contribution in the H$\alpha$ line showed that it was not possible to make a high precision telluric correction. Therefore, we concluded that utilizing the spectra we have accessible in our archival data set we cannot perform such a correction at the desired level.  We do acknowledge that leaving the $EW$ measures uncorrected for telluric contribution leaves the systematic effect of telluric absorption in our results. However, we choose not to increase the uncertainty of our results by applying a correction factor that can produce errors as large (if not larger) than the underlying contribution we are trying to correct for. Therefore, our estimate for the uncertainty associated with the $EW$ measure is fixed at the 3\% level.  

When more than one observation
was present for a given target star, we have also calculated the mean
uncertainty
\begin{equation}
\label{sigma_bar}
\overline{\sigma} = \frac{\sum^N_{i=1}\sigma_i}{N},
\end{equation}
where $\sigma_i$ is the uncertainty of the $i_{th}$ data point, and
$N$ is the total number of spectra of a given target star. Similarly, the 
standard deviation of the sample was calculated using
\begin{equation}
\label{standard_deviation}
s = \sqrt{\frac{1}{N-1}\sum^N_{i=1}\left(EW_i - \overline{EW}\right)^2}.
\end{equation}
$N - 1$ in Equation 2 corresponds to the degree of freedom used for the $F$ test. Having obtained ${\overline{\sigma}}$ and $s$ allowed us to calculate
a simple observable to assess the variability of a given
target using a ratio of the form
\begin{equation}
\label{ratio}
F = \frac{s^2}{\overline{\sigma}^2}.
\end{equation}	

\noindent This ratio along with the number of spectra allows a quantitative assessment to be assigned to the 
variability of a particular target in terms of the amount of change in
the $EW$ of  H$\alpha$. Therefore, the emphasis of this work is subtly
different than the focus of many studies in the literature that have
focused on obtaining the period of the variability.

We have calculated the F ratio and corresponding confidence level for all our targets. The individual 
values are listed in Table~\ref{table_summary}. The confidence level, $C$, takes into account the number of spectra available for each target. Therefore this column should be used assess the variability of a particular target. For example, for HR 193, the first star in Table~\ref{table_summary}, the 
$F$ ratio is 1.73 and $C$ is 0.86. This means that this source is variable at the 86\% confidence level, or in other words this is the confidence level at which we conclude that the sample variance based on the entire set of spectra for HR 193 and the variance based on the mean uncertainty are different.

The program stars 
were also investigated for obvious trends in the H$\alpha$ $EW$, such as increasing or 
decreasing over the duration of this study. In order to quantitatively assess a trend, we fit a line to each set of H$\alpha$ $EW$s for a particular target.  We calculated the slope and error from line fitting for each line. Since we are only interested in assessing the trend and not in the numerical value of the slope, observational errors were not included in the slope fitting.  Also, for this analysis we did not include any stars for which only two observations were available. Table~\ref{table_trends} lists the slope and the slope divided by the error for each target.  Recall that since we have adopted the convention that emission corresponds to a negative $EW$, negative slopes correspond to an increase in H$\alpha$ $EW$ and vice versa.  In order to help the reader assess these numbers, in the final column of Table~\ref{table_trends}, a trend of gain (G) or loss (L) in H$\alpha$ $EW$ is assigned to targets with a 3 $\sigma$ detection or greater and are variable with a confidence of greater than 90\%.

\section{Results}

The $C$ value for each program star is plotted in Figure~\ref{mar_his} and the frequency with which each value occurs is shown in the inverted histogram below the abscissa. The value of $C$ ranges from a minimum 
of 0 to a maximum of greater than 99\%.
Of the 56 targets analyzed, 77\% of them were categorized as variable with a $C$ of 90\% or greater. Only 11\% of the program stars were variable with a confidence of less than 50\%. Within this group of six targets, 4 had only 2 or 3 spectra obtained resulting in the low value of $C$.  The other two stars, $\psi$ Per (HR 1087) and 48 Per (HR 1273) have 17 and 18 spectra with $C$ of 34\% and 0\%, respectively. This points to the global stability of the H$\alpha$ emitting region for these two systems.    

Figure~\ref{cas} shows the plot of $EW$ versus time for the early Be star, $\gamma$ Cas (HR 264).  The application of the F test reveals that this system is variable at the confidence level of 94\%. We note that this star shows more moderate variations in the H$\alpha$ $EW$ ranging from a minimum of -2.88 nm to a maximum of -3.39 nm over the time it was monitored in our study.  Despite these moderate variations in $EW$, the results from the line fitting analysis (shown Table~\ref{table_trends}) also allow us to designate this star as showing a definite H$\alpha$ $EW$ increasing trend (G) at the 12 $\sigma$ level. This illustrates the point that the variability criteria using this statistical technique can be adjusted as needed, depending on the needs of a specific observing program.  

$\beta$ Lyr (HR 7106) is an example of a star that we find is variable with 
$C>99\%$. The plot of $EW$ versus time for this star is shown in Figure~\ref{lyr}.  The H$\alpha$ $EW$ ranges from a minimum of -1.15 nm to a maximum of -2.68 nm over the time it was monitored in this investigation.  For this star, although it is certainly variable, we do not find an obvious trend in H$\alpha$ $EW$ (see Table~\ref{table_trends}). This designation does not necessarily mean that there is no periodicity; it simply means that there is no monotonically increasing or decreasing trend. Perhaps this should be expected based on the close binary nature of this source \citep{zha08, sch09}. This star is a known interacting binary with an orbital period of 12.9421 days \citep{kre04}. The random (perhaps periodic) nature of this variability can be seen in Figure~\ref{lyr}. 

Figure~\ref{ori} shows the plot of $EW$ for the star $\omega$ Ori (HR 1934) and is another example of a variable star with $C>99\%$. Overall, it may appear that there is a trend towards less negative numbers corresponding to a decrease in overall H$\alpha$ emission.  However, we were not able to make this claim at the 3 $\sigma$ level and have chosen instead, to present the numbers in Table~\ref{table_trends} and allow the reader to make their own assessment.  This may be an example where more observations are required to determine whether this is a monotonic decreasing trend or whether this is periodic or random behavior.

Figures~\ref{cam} and~\ref{cyg}, for the stars BK Cam (HR 985) and 28 Cyg (HR 7708), respectively, demonstrate further examples of stars that are variable with a confidence level of greater than 99\%. Notice that the plot of $EW$ versus time for BK Cam clearly shows an increasing trend in H$\alpha$ $EW$ over the duration of this study, whereas the plot for 28 Cyg shows a decreasing trend. The values presented in Table~\ref{table_trends} support this interpretation. 

\section{Discussion}

We have a number of stars in common with other studies of Be star variability including 
studies of photometric variability and line profile variations. While this 
study
focuses specifically on changes in the H$\alpha$ equivalent width, it is interesting to 
compare our results with properties others have found for the 
same stars. Rather than discuss every star we have in common with previous investigations, we have 
chosen a selection of some of the more interesting cases and cases that highlight 
similarities or differences between our study and other work in the
literature. This discussion 
is provided to allow the reader to assess the value of this simple, but 
illuminating technique.

One of our program stars, $\omega$ CMa (HR 2749) has been previously studied by
\citet{riv03}, \citet{mai03}, and \citet{stef99} and its' line profile variability is well known. This star has been found to have a period of $\sim$ 1.5 days \citep{ste99}. 
For this particular star we only have
three observations, and in combination with the fact that this star is variable on 
a relatively short time scale our ratio will be susceptible to sampling 
limitations. Not surprisingly our results fail to establish variability of this source, which is also reflected in a $C$ value of only 33\%. We also note that it is possible to have changes in the shape of the H$\alpha$ 
profile without affecting the value of the H$\alpha$ equivalent width as long as 
increases or decreases in emission or absorption cancel out. For example, there could be a 
significant change in the V/R (violet to red) ratio where the reduction in $EW$ due to a weaker V component is compensated by an increase in R, or vice versa. However, other
program stars that vary on short time
scales were captured by our technique. For example, $\phi$ Per (HR 496) is a known short period
variable \citep{hub98} and our study clearly identifies this star as variable with a confidence of $>99\%$. Furthermore, our line fitting technique (see Table~\ref{table_trends}) indicates that this system is undergoing a reduction in H$\alpha$ emission.
Neveretheless, $\omega$ CMa is a good example to illustrate the subtle difference in this technique compared with other studies. This tool is designed
to capture changes in the H$\alpha$ equivalent width. The H$\alpha$ emitting region
typically extends several stellar radii from the central star \citep{tyc06}. 
Large changes in H$\alpha$ equivalent
width will occur with variations in disk density, changes in optical depths that affect emission, or
a change in the distribution of disk material. As such, our application of the F test is designed to identify systems that may be experiencing significant changes in their density and thermal structure. 

For the Be star $\upsilon$ Cyg (HR 8146) we find that it is variable with a confidence of $>99\%$. This star was 
previously 
identified as having long period variability by \citet{hub98} and also was 
noted 
as variable by \citet{per02}. \citet{hub98} mention that this star exhibited a strong increase in brightness followed by a very slow decline
over 400 days. \citet{per02} correlate the amplitude of the short 
period variations with possible disk growth and loss suggested by changes in 
brightness over a longer term. Our technique agrees with and reflects 
the nature of 
the variability for this star.

Large changes in H$\alpha$ equivalent width have been noted in the literature for 66 Oph (HR 6712). \citet{pet89,pet92,pet94,han96} and \citet{hub98} all report large changes, with
Hipparcos also showing 3 large outbursts \citep{hub98}. We find that his star is variable 
with $C>99\%$. Our result is in clear agreement with these studies. Furthermore, we find that during the time of our observations that the H$\alpha$ $EW$ for 66 Oph was undergoing a decreasing trend. Similarly 
$\omega$ Ori (HR 1934) has been noted to exhibit recurrent outbursts \citep{pet96}. We find that this star is variable at $>99\%$. As expected, our technique captures the change expected due to these large outbursts.
  
\citet{hub98} find that larger amplitudes of light variability using Hipparcos 
photometry are found more often in the early Be stars (33\% of B0-B2e) compared with late types (10\% B7-B9e). \citet{hub98} do have a larger set of stars 
but a comparison of their figure 1 with our Figure~\ref{type} shows the number 
distribution for both studies as of function of spectral type are quite 
similar. Their finding is consistent
with our statistics especially considering that we expect our technique to be most sensitive to significant changes in the disk density. We find 55\% of our sample is variable with confidence
of 99\% or greater. Within this group of variable stars, 45\% are early Be stars (B0-B2) while 29\% are late-type (B7-B9).
 
\begin{figure}
\plotone{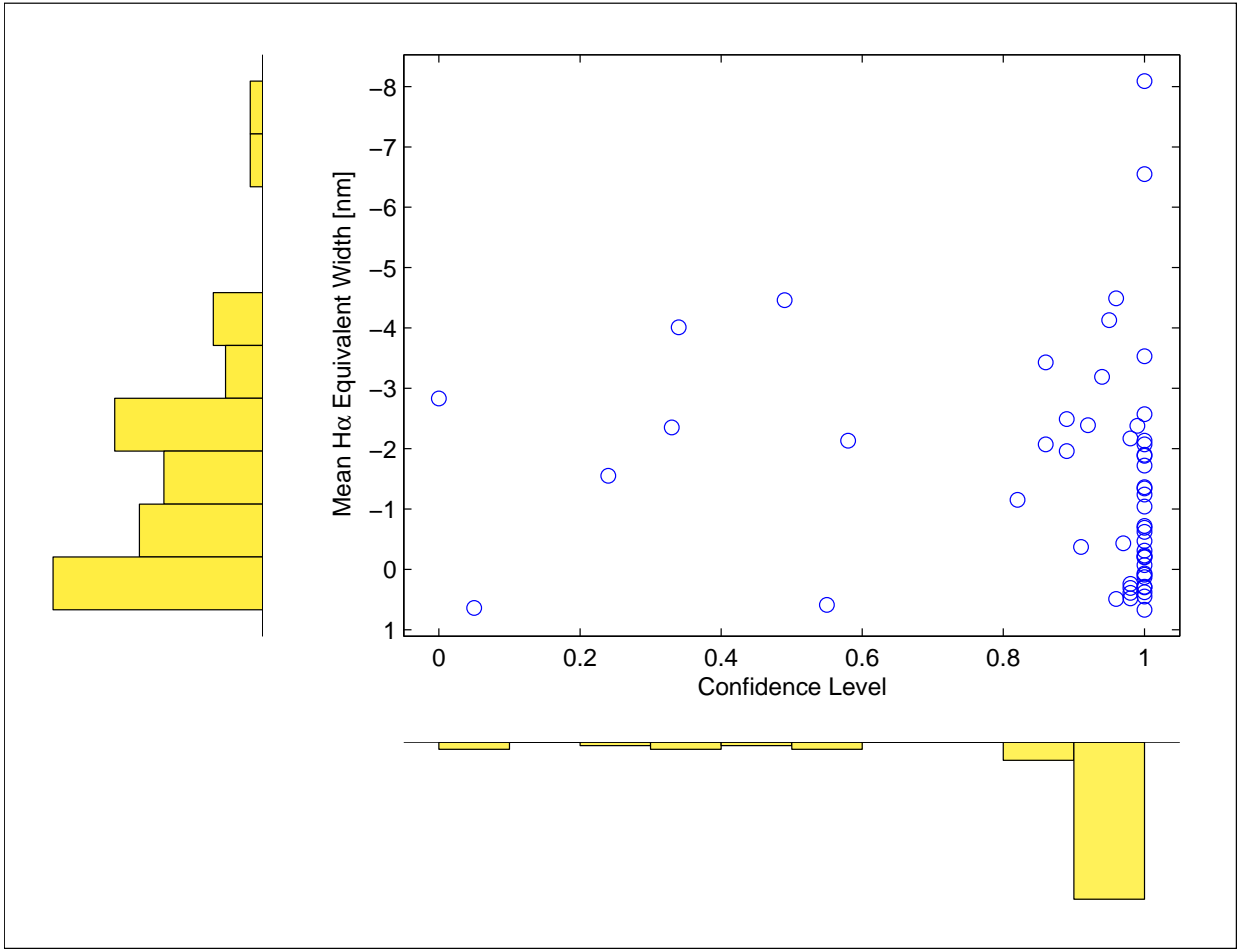}
\caption{The correlation of ${\overline{EW}}$ versus $C$ for the program stars. The histograms show the number of program stars binned with respect to ${\overline{EW}}$ and $C$ on the vertical and horizontal axes, respectively.
}
\label{mar_his}
\end{figure}
\begin{figure}
\plotone{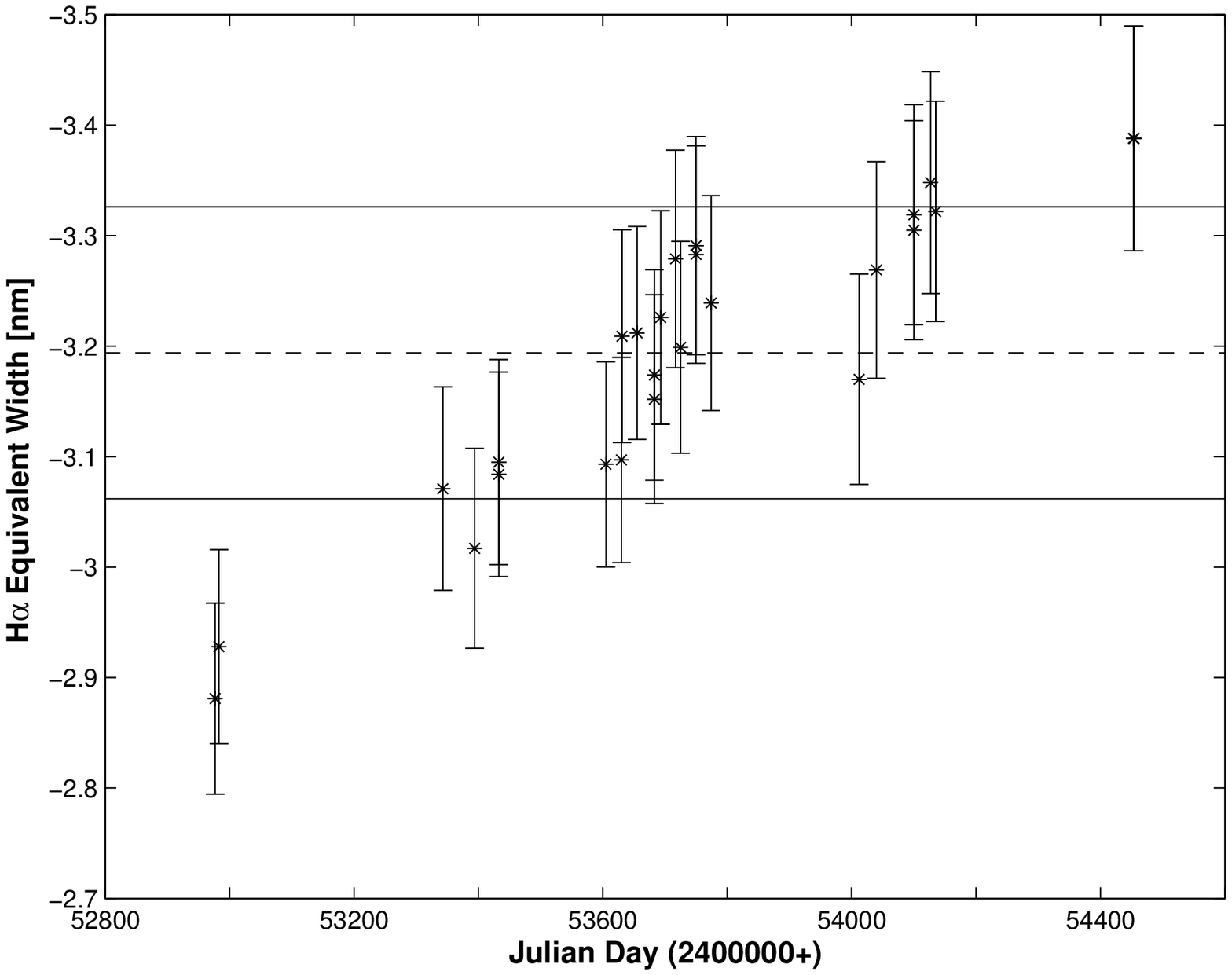}
\caption{Same as Figure~\ref{example} except for the star $\gamma$ Cas (HR
264).
}
\label{cas}
\end{figure}
\begin{figure}
\plotone{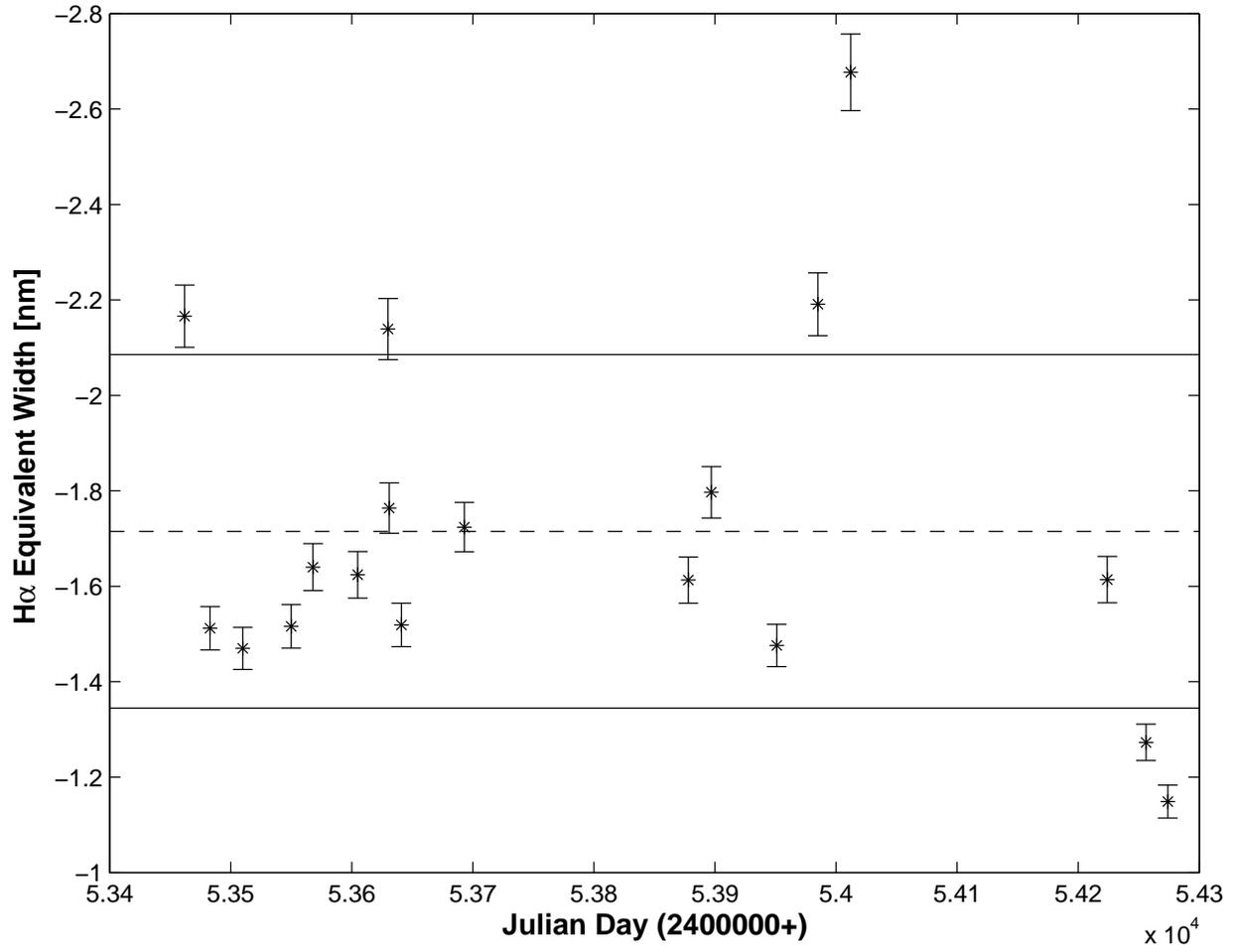}
\caption{Same as Figure~\ref{example} except for the star $\beta$ Lyr (HR 7106).
}
\label{lyr}
\end{figure}
\begin{figure}
\plotone{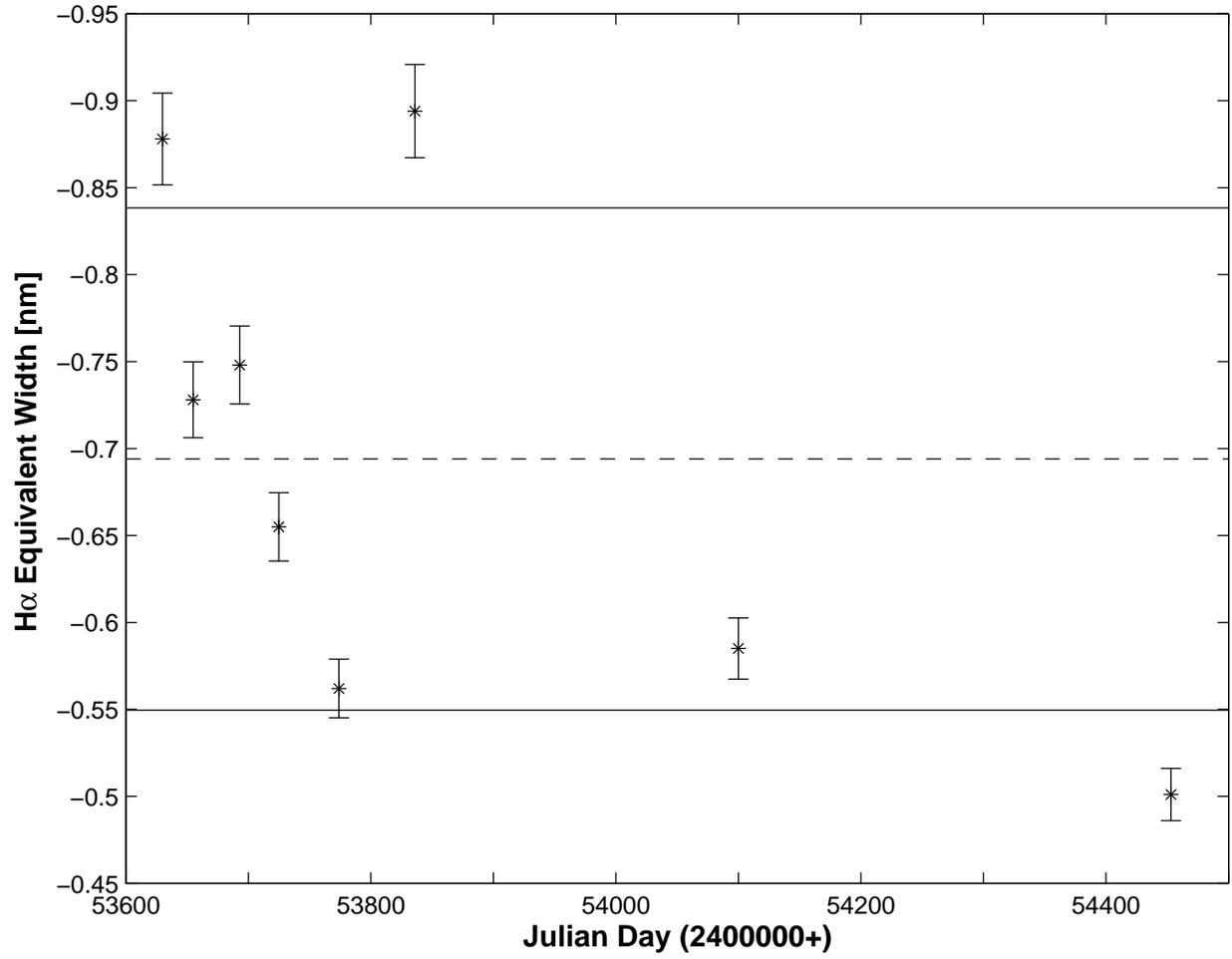}
\caption{Same as Figure~\ref{example} except for the star $\omega$ Ori 
(HR 1934).
}
\label{ori}
\end{figure}
\begin{figure}
\plotone{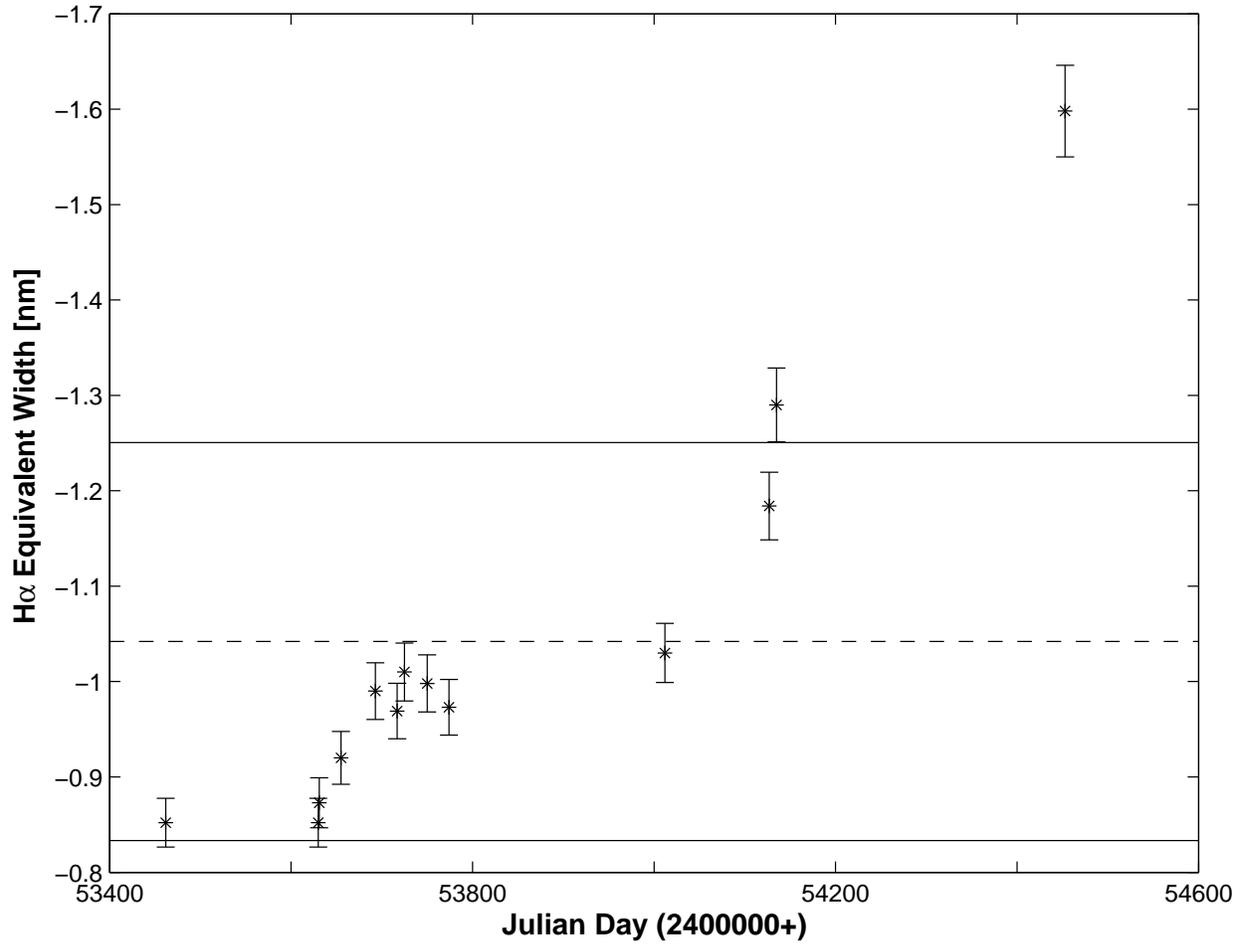}
\caption{Same as Figure~\ref{example} except for the star BK Cam (HR 985).
}
\label{cam}
\end{figure}
\begin{figure}
\plotone{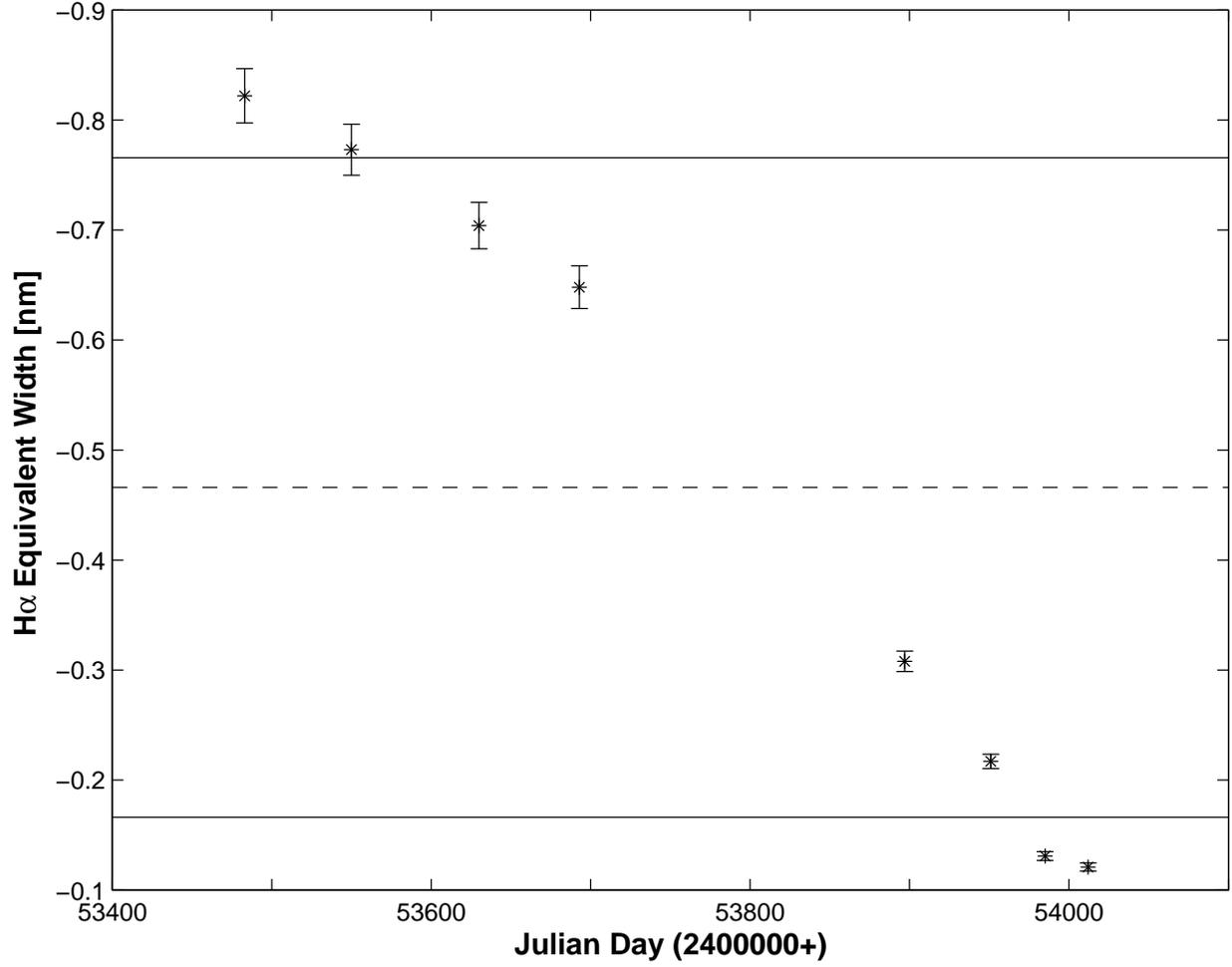}
\caption{Same as Figure~\ref{example} except for the star 28 Cyg (HR 7708).
}
\label{cyg}
\end{figure}

We also considered whether or not the application of the F test may be biased toward finding variability with a high confidence level for stars with particularly strong H$\alpha$ emission.  Figure~\ref{mar_his} (discussed above) also shows the correlation of ${\overline{EW}}$ versus $C$.  It is clear that stars represented by variability with the highest confidence of greater than 90\% exhibit a substantial range in the strength of H$\alpha$ $EW$. In fact, as shown in Figure~\ref{mar_his} the highest values of $C$ have the greatest proportion of stars with weak emission. This supports the view that this simple statistical technique can be used to effectively to find Be stars at particular stages.

\section{Summary}

We have investigated, a simple but quantitative way to determine whether a 
Be star is variable based on $EW$ measurements of the H$\alpha$ emission line. Since the H$\alpha$ spectral line is often the most prominent feature in Be star spectra, the H$\alpha$ spectra are often easy to obtain through new observations or are often available in the literature. Our application of the F test based on
${\overline{\sigma}}$ and $s$, provides a simple quantitative assessment about 
the change in the $EW$ of H$\alpha$ for an object.
This technique can be used to find Be stars in particular 
stages of variability or alternatively can be used to determine when a star/disk system is 
changing 
and needs to be monitored closely. The specific connection between photometric variability and changes in 
H$\alpha$ equivalent 
width remains unclear and improvements in understanding Be star variability is key to the development and testing
of dynamical models \citep{hub98}. We should also mention that our measurements do not differentiate from line emission changes or continuum changes and either of these could result in changes in $EW$ measure. 

A prerequisite to the development of successful dynamical models will be timely observations that 
adequately sample the disk-loss, and disk-growth events of classical Be stars. \citet{mcs09} 
studied Be stars in southern open clusters and found 12 new Be stars that 
transitioned between
B and Be phases, clear evidence that it is possible to find reasonably sized samples to
investigate particular evolutionary stages. These 
transition phases were 
recently investigated using spectropolarimetric data for the 
Be stars, $\pi$~Aquarii and 60~Cygni by \citet{wis10}. $\pi$~Aqr (HR 8539) is one of our program stars that we find variable. We also find a trend in decreasing H$\alpha$ $EW$ during our study (see Table~\ref{table_trends}). Interestingly, \citet{wis10} find that during one of 
the disk loss phases
there were two extended outburst events.  These events seemed to halt the disk 
loss phase for some time. 
We have a significant data set for $\pi$ Aqr, and the most recent spectra 
we obtained, suggests 
that the H$\alpha$ $EW$ may be starting to increase again.  
It is clear that observations acquired at particular stages are key to improving our understanding of Be stars and their intrinsic variability.  

\acknowledgments

We thank the Lowell Observatory for the telescope time used to
obtain the H$\alpha$ line spectra presented in this work. We are grateful
to the anonymous referee whose thorough review helped to improve the paper and to Dietrich Baade for his insightful comments. This research was 
supported in part
by NSERC, the National Sciences and Engineering Research Council of Canada.
C. T. acknowledges, with thanks, grant support from the Central Michigan 
University. This research has made use of the SIMBAD database, operated at CDS, Strasbourg, France.

{\it Facilities:} \facility{Hall}.

\clearpage
\begin{deluxetable}{lccccccc}
\tablecaption{Program Stars and their F Value
\label{table_summary}}
\tablehead{
\colhead{HR \#}	&	{HD \#}	&	{Name}	&	{Spectral Type}	&\# &	{$\overline{EW}$} [nm]	& F value & $C$ \tablenotemark{a}}
\startdata
193	&	4180	&	{\em o} Cas	&	B5IIIe	&18&	-3.43	& 1.73&0.86\\
264	&	5394	&	$\gamma$ Cas	&	B0IVe	&26&	-3.19	& 1.90&0.94\\
335	&	6811	&	$\phi$ And	&	B7Ve	&13&	0.13	& 35.40&$>$0.99\\
496	&	10516	&	$\phi$ Per	&	B2Vpe	&20&	-3.53	& 7.19& $>$0.99\\
	&	11606	&	V777 Cas	&	B2Vne	&2&	-1.96	& 30.61& 0.89 \\
936	&	19356	&	$\beta$ Per	&	B8V	&13&	0.67	& 11.39&$>$0.99\\
	&	19243	&	V801 Cas	&	B1Ve	&3& -4.49	& 23.81&0.96\\
985	&	20336	&	BK Cam	&	B2.5Vne	&13&	-1.04	&44.59&$>$0.99\\
1087	&	22192	&	$\psi$ Per	&	B5Ve&17	&	-4.01	& 0.81&0.34\\
1142	&23302	&17 Tau	&	B6IIIe	&11& 0.39&4.26&0.98\\
1156	&	23480	&	23 Tau	&	B6IVe	&9&	0.07	&486.89 &$>$0.99\\
1165	&	23630	&	$\eta$ Tau	&	B7IIIe&11	&-0.31	&12.24 &$>$0.99\\
1180	&	23862	&	28 Tau	&	B8IVevar&12	&	-2.07	&144.37 &$>$0.99\\
1209	&	24534	&	X Per	&	O9.5pe	&4&	-2.38	& 23.68&0.99\\
1261	&	25642	&	47 Per	&	A0IVn	&2&	0.64	& 0.01& 0.05\\
1273	&	25940	&	48 Per	&	B3Ve	&18&	-2.83	& 0.26&0.00\\
1508	&	30076	&	56 Eri	&	B2Ve	&3&	-4.13	& 17.73&0.95\\
1605	&	31964	&	$\epsilon$ Aur	&	A8Iab	&6&	0.09	& 84.16&$>$0.99\\
1622	&	32343	&	11 Cam	&	B2.5Ve	&9&	-2.39	& 2.90&0.92\\
1660	&	32991	&	105 Tau	&	B3Ve	&2&	-4.46	& 0.94& 0.49\\
1789	&	35439	&	25 Ori	&	B1Ve	&6&	-1.34& 34.42&$>$0.99\\
1910	&	37202	&	$\zeta$ Tau	&	B2IVe&38	&	-1.90	& 26.55&$>$0.99\\
1934	&	37490	&	$\omega$ Ori	&	B2IIIe&8	&	-0.69	& 48.13&$>$0.99\\
2148	&	41511	&	17 Lep	&	Apsh	&4&	0.45	& 121.60&$>$0.99\\
2343	&	45542	&	$\nu$. Gem	&	B6IIIe&13	&	-0.07	& 57.93&$>$0.99\\
2538	&	50013	&	$\kappa$ CMa	&	B1.5IVe	&4&	-2.17	& 18.41&0.98\\
2749	&	56139	&	$\omega$ CMa	&	B2IV - Ve	&3&	-2.35	& 0.50&0.33\\
2845	&	58715	&	$\beta$ Cmi	&	B8Ve	&22&	-0.20	& 18.34&$>$0.99\\
3034	&	63462	&	{\em o} Pup	&	B1IV:nne&2	&-1.55	& 0.16 & 0.24\\
4696	&	107348	&	5 Crv	&	B8V	&9&	-0.37	& 2.69&0.91\\
4787	&	109387	&	$\kappa$ Dra	&	B6IIIpe&21	&-2.13		& 1.10&0.58\\
	&	141569	&		&	B9.5e	&6&	0.30	&22.22 &$>$0.99\\
5938	&	142926	&	4 Her	&	B9pe	&6&	-0.21	& 82.39&$>$0.99\\
5941	&	142983	&	48 Lib	&	B3Ia/Iab&13	&	-2.49	& 2.07&0.89\\
5953	&	143275	&	$\delta$ Sco	&	B0.2IVe&37	&	-1.88	& 126.29&$>$0.99\\
6118	&	148184	&	$\chi$ Oph	&	B2Vne	&20&	-6.55	& 11.51&$>$0.99\\
6397	&	155806	&	V1075 Sco	&	O8Ve	&4&	-0.43	& 15.34&0.97\\
6519	&	158643	&	51 Oph	&	A0V 	&7&	0.31	& 7.04&0.98\\
6712	&	164284	&	66 Oph	&	B2Ve	&15&	-0.72	& 125.11&$>$0.99\\
6779	&	166014	&	{ \em o} Her	&	B9.5V	&6&	0.59	& 1.13&0.55\\
7040	&	173370	&	4 Aql	&	B9V	&5&	0.24	& 10.29&0.98\\
7106	&	174638	&	$\beta$ Lyr	&	B7Ve+...&18	&	-1.72	& 51.82&$>$0.99\\
	&	179218	&	MWC 614	&	B9e	&3&	-0.38	& 801.34&$>$0.99\\
7708	&	191610	&	28 Cyg	&	B2.5Ve	&8&	-0.47	& 460.73&$>$0.99\\
7739	&	192685	&	QR Vul	&	B3Ve	&5&	0.29	& 638.97&$>$0.99\\
7763	&	193237	&	P Cyg\tablenotemark{b} 	&	B2pe	&62&	-8.09		&3.60 &$>$0.99\\
7836	&	195325	&	1 Del	&	A1she...&5	&	0.49	& 7.15&0.96\\
7963	&	198183	&	$\lambda$ Cyg	&	B5Ve&6	&	0.48	& 8.54&0.98\\
8047	&	200120	&	59 Cyg	&	B1.5Vnne&7	&	-1.24	& 12.10&$>$0.99\\
8146	&	202904	&	$\upsilon$ Cyg	&	B2Vne&18	&	-2.57	&5.62 &$>$0.99\\
8260	&	205637	&	$\epsilon$ Cap	&	B3V:p&6	&	-0.23	& 153.86&$>$0.99\\
	&	206773	&		&	B0Vpe	&2&	-1.15	& 11.47& 0.82\\
8402	&	209409	&	{\em o} Aqr	&	B8IVe 	&12&	-2.07	&1.97 &0.86\\
8520	&	212076	&	31 Peg	&	B2IV - Ve&11	&	-2.13	& 39.00&$>$0.99\\
8539	&	212571	&	$\pi$ Aqr	&	B1Ve&11	&	-0.62	& 146.25&$>$0.99\\
8773	&	217891	&	$\beta$ Psc	&	B6Ve&15	&	-1.36	&27.42 &$>$0.99\\
\enddata
\tablenotetext{a}{Confidence level of the detected variability based on the ratio of the variances.}
\tablenotetext{b}{This star is a Luminous Blue Variable, LBV.}
\end{deluxetable}

\clearpage
\clearpage
\begin{deluxetable}{lccccc}
\tablecaption{Variability Trends \label{table_trends}}
\tablehead{
\colhead{HR \#}	&	{HD \#}	&	{Name}	&	{Slope}	& {Slope/Uncertainty} &	{Trend}}
\startdata
193	&	4180	&	{\em o} Cas	&	-0.000441&-8.076&-\\
264	&	5394	&	$\gamma$ Cas	& -0.000329 &-11.592&G\\		
335	&	6811	&	$\phi$ And	&	1.206e-05 &0.442&-\\
496	&	10516	&	$\phi$ Per	&	0.000709 &4.072&L\\
936	&	19356	&	$\beta$ Per	& .00010	& 1.369&-\\
	&	19243	&	V801 Cas	&	0.001634 & 9.682&L\\
985	&	20336	&	BK Cam	        &	-0.000723 & -10.026&G\\
1087	&	22192	&	$\psi$ Per	&	-0.000361 & -10.00&-\\
1142	&	23302	&	17 Tau	        &	-6.98e-05& -4.525&G\\
1156	&	23480	&	23 Tau	&	9.64e-05& 1.805&-\\
1165	&	23630	&	$\eta$ Tau	&-2.32e-05 & -0.956&-\\
1180	&	23862	&	28 Tau	&	0.0027 & 9.795&L\\
1209	&	24534	&	X Per	&	-0.000959 & -3.842&G\\
1273	&	25940	&	48 Per	&	-7.03e-05 & -2.207&-\\
1508	&	30076	&	56 Eri	&	-0.0013 & -3.350&G\\
1605	&	31964	&	$\epsilon$ Aur	&-6.21e-05 & -1.157&-\\
1622	&	32343	&	11 Cam	&	0.000353 & 4.091&L\\
1789	&	35439	&	25 Ori	&	-0.000457 & -1.633&-\\
1910	&	37202	&	$\zeta$ Tau	&	6.50e-05 & 0.579&-\\
1934	&	37490	&	$\omega$ Ori	&	0.000335&  2.125&-\\
2148	&	41511	&	17 Lep	&	0.000390& 1.111&-\\
2343	&	45542	&	$\nu$. Gem	&	-3.712e-05 & 2.722&-\\
2538	&	50013	&	$\kappa$ CMa	&	0.000272 & 0.069&-\\
2749	&	56139	&	$\omega$ CMa	& 8.16e-05 & 0.473&-\\
2845	&	58715	&	$\beta$ Cmi	&	-1.304e-05 & -0.804&-\\
4696	&	107348	&	5 Crv	& -3.22e-05	& -1.536&-\\
4787	&	109387	&	$\kappa$ Dra	&	0.000159 & 4.602&-\\
	&	141569	&		&	0.000849 & 2.790&-\\
5938	&	142926	&	4 Her	&	0.000160 & 3.421&L\\
5941	&	142983	&	48 Lib	&	-0.000278 & -5.844&-\\
5953	&	143275	&	$\delta$ Sco	&	0.0018& 10.650&L\\
6118	&	148184	&	$\chi$ Oph	&	0.0012 & 2.35&-\\
6397	&	155806	&	V1075 Sco	&	-0.000157 & -1.318&-\\
6519	&	158643	&	51 Oph	&	4.263e-05 & 1.042&-\\
6712	&	164284	&	66 Oph	&	0.000696 & 7.471&L\\
6779	&	166014	&	{ \em o} Her	&	2.304e-05 & 0.801&-\\
7040	&	173370	&	4 Aql	&	6.832e-05 & 3.88&L\\
7106	&	174638	&	$\beta$ Lyr	&	0.000187 & 0.560&-\\
	&	179218	&	MWC 614	&	-0.0019 & -16.310&G\\
7708	&	191610	&	28 Cyg	&	0.0014 & 18.74&L\\
7739	&	192685	&	QR Vul	&	0.000257 & 0.573&-\\
7763	&	193237	&	P Cyg 	&	-0.000641 & -4.154&- \tablenotemark{a}\\
7836	&	195325	&	1 Del	& -5.03e-05& -0.681&-\\
7963	&	198183	&	$\lambda$ Cyg	&-6.55e-5&-0.791&-\\
8047	&	200120	&	59 Cyg	& -0.00027 & -2.146&-\\
8146	&	202904	&	$\upsilon$ Cyg	&0.00038& 2.876&-\\
8260	&	205637	&	$\epsilon$ Cap	&-0.0002& 0.776&-\\
8402	&	209409	&	{\em o} Aqr	&3.62e-05& 0.385&-\\
8520	&	212076	&	31 Peg	&	-0.0013&-5.54&G\\
8539	&	212571	&	$\pi$ Aqr	&	0.00076& 5.36&L\\
8773	&	217891	&	$\beta$ Psc	&0.0008 & 12.4&G\\
\enddata
\tablenotetext{a}{We chose not to designate this star as increasing since we obtained a large number
of spectra on 2 nights which may have biased the slope.}

\end{deluxetable}

\clearpage
\end{document}